\documentclass[journal]{IEEEtran}
\IEEEoverridecommandlockouts
\usepackage[utf8]{inputenc}
\usepackage{amsmath}
\usepackage{cite}
\usepackage{multicol}
\usepackage{graphicx}
\usepackage{enumitem}
\usepackage{array}
\usepackage{multirow}
\usepackage{graphicx}
\usepackage{enumitem}
\usepackage{color}
\usepackage{nomencl}
\usepackage{etoolbox}
\usepackage{amssymb}
\usepackage{comment}
\usepackage{mathtools}
\usepackage{bbold}
\usepackage{yfonts}
\usepackage{soul}
\usepackage{hyperref}
\usepackage{dsfont}
\usepackage[acronym]{glossaries}
\usepackage{nomencl}
\usepackage{subfiles}
\usepackage[utf8]{inputenc}
\usepackage{mathtools}
\usepackage{amsmath}
\usepackage{tikz}
\usepackage{algorithm}
\usepackage{algpseudocode}
\usepackage{fnpct}
\ifCLASSOPTIONcompsoc \usepackage[caption=false,font=normalsize,labelfon
t=sf,textfont=sf]{subfig}
\else
\usepackage[caption=false,font=footnotesize]{subfig}
\fi
\usepackage{tikz}
\usetikzlibrary{decorations.pathreplacing}
\usetikzlibrary{arrows,shapes,positioning}
\usetikzlibrary{patterns}
\usepackage{pgfplots}
\def\BibTeX{{\rm B\kern-.05em{\sc i\kern-.025em b}\kern-.08em
    T\kern-.1667em\lower.7ex\hbox{E}\kern-.125emX}}
\begin{document}

\title{Grid-aware Scheduling and Control of Electric Vehicle Charging Stations for Dispatching {Active Distribution Networks}. Part-II: Intra-day and Experimental Validation}
\author{
    \thanks{This project has received funding in the framework of the joint programming initiative ERA-Net Smart Energy Systems’ focus initiatives Smart Grids Plus and Integrated, Regional Energy Systems, with support from the European Union’s Horizon 2020 research and innovation program under grant agreements No 646039 and 775970. (\textit{Corresponding author: Rahul K. Gupta.)}}
    \IEEEauthorblockN{
    Rahul K. Gupta$^*$\thanks{$^*$School of Electrical and Computer Engineering, Georgia Institute of Technology, Atlanta, 30308, USA (e-mail: rahul.gupta@gatech.edu).}, Sherif Fahmy$^\S$, Max Chevron$^\S$, Enea Figini$^\S$, Mario Paolone$^\S$\thanks{$^\S$Distributed Electrical Systems Laboratory, École Polytechnique Fédérale de Lausanne, 1015 Lausanne, Switzerland (e-mail: \{sherif.fahmy, max.chevron, enea.figini, mario.paolone\}@epfl.ch).}\\}
    }
\makeatletter
\patchcmd{\@maketitle}
  {\addvspace{0.5\baselineskip}\egroup}
  {\addvspace{-1.5\baselineskip}\egroup}
  {}
  {}
\maketitle
%%%%%%
\begin{abstract}
In Part-I, we presented an optimal day-ahead scheduling scheme for dispatching active distribution networks accounting for the flexibility {provided by} electric vehicle charging stations (EVCSs) and other controllable resources such as battery energy storage systems (BESSs). Part-II presents the intra-day control layer for tracking the dispatch plan computed from the day-ahead scheduling stage. The control problem is formulated as model predictive control (MPC) with an objective to track the dispatch plan setpoint every 5~minutes, while actuated every 30~seconds. MPC accounts for the uncertainty of the power injections from stochastic resources (such as demand and generation from photovoltaic -- PV plants) by short-term forecasts. MPC also accounts for the grid's operational constraints (i.e., the limits on the nodal voltages and the line power-flows) by a linearized optimal power flow (LOPF) model based on the power-flow sensitivity coefficients, and for the operational constraints of the controllable resources (i.e., BESSs and EVCSs). The proposed framework is experimentally validated on a real-life ADN at the EPFL's Distributed Electrical Systems Laboratory {and} is composed of a medium voltage (MV) bus connected to three low voltage distribution networks. It hosts two controllable EVCSs (172~kWp and 32~kWp), multiple PV plants (aggregated generation of 42~kWp), uncontrollable demand from office buildings (20~kWp), and two controllable BESSs (150kW/300kWh and 25kW/25kWh). 
\end{abstract}
%%%%%%%%%%%%%%%%%%
\section{Introduction}
\subsection{{Background}}
Real-time control of Electric Vehicle Charging Stations (EVCSs) has been widely {advocated} for supporting {power} grid operations such as primary and secondary frequency control \cite{zhong2014coordinated, janjic2017commercial}, voltage support \cite{bansal2014plug}, three-phase demand and voltage balancing in \cite{weckx2015load, fahmy2020grid}, bidirectional power/energy balancing in form of short-term storage via Vehicle to Grid (V2G) control \cite{liu2014vehicle, hernandez2018primary}, etc. 
This work is interested in using EVCS flexibility for dispatching services which consist of tracking a pre-defined power schedule at the grid connection point (GCP) of an active distribution network (ADN). 
The advantages of such dispatching schemes {refer to supporting} the bulk transmission system power and energy imbalances at the local scale and solving local distribution grid operational issues {\cite{nottrott2013energy, reihani2016energy, bozorg2018influencing}.} 
In particular, we refer to the Part-I paper, where a dispatching framework using EVCS flexibility was introduced and an optimal day-ahead scheduling scheme (to optimize a dispatch plan) was presented while accounting for the flexibility from EVCSs and other controllable resources such as battery energy storage system (BESS). However, due to uncertainty on the day-ahead generation and demand, it is necessary to have an intra-day real-time controller that can track the day-ahead schedule accurately, using flexible resources by regulating their power injections.
In this context, the objective of this {Part-II paper} is to develop a real-time control algorithm for tracking dispatch plan at the ADN's GCP by regulating power injections from EVCSs and BESSs. Aside from the dispatch tracking objective, the controller also maximizes EV user satisfaction while minimizing the battery wear of every EV and fairly allocating the power {required to charge multiple EVs}. 

\subsection{{Related Work}}
Different control formulations have been proposed in the literature for EVCS control. For example, in \cite{hu2016multi}, an agent-based online optimization of EVCS operation was used for the grid congestion management. A heuristic controller was proposed in \cite{qi2014hierarchical, imran2020heuristic} {for efficient energy management,} where the EVCS power was regulated using measurements of the power injection and the state-of-charge. In \cite{al2021bidirectional, wang2012dynamic}, a lookup-table-based controller was deployed, in which the control setpoints {follow a pre-scheduled profile.} In \cite{tuchnitz2021development, dabbaghjamanesh2020reinforcement}, a reinforcement learning-based data-driven controller was proposed. All these schemes did not incorporate any time-dependent constraints {that due to the} inherent energy storage characteristics of EVs {must be considered in the control problem.}
Model predictive control (MPC)-based schemes are widely used for EVCS control \cite{bansal2014plug, di2014electric, tang2016model, wang2022mpc} to account for the time-dependent constraints of EV storage capacity. In \cite{zhou2022integrated}, MPC was used for energy management of the EV demand with an objective to maximize EV user satisfaction. In \cite{van2021peak, li2021coordinating}, MPC-based control was used for peak-shaving and load leveling.  In this context, we formulate our real-time controller as an MPC problem with an objective to track the dispatch plan at each timestep (5~minutes). The MPC is solved every 30~seconds, accounting for the dispatch error that occurred during previous time steps, and eventually nullifying it by the end of the current 5-minutes dispatch horizon. 
 
Furthermore, most of the existing literature on EVCS control is limited to numerical validation and lacks practical relevance with respect to real-life implementation. Very few works have performed {extensive} experimental real-life validation of the EVCS control due to the potential technical difficulties associated with it. In this context, this paper also performs an experimental validation of the proposed MPC-based control of EVCS on a real-life ADN. The experimental validation is performed for multiple continuous days to demonstrate that the EVCS flexibility can be reliably used for dispatching ADNs. 

\subsection{{Proposed work and contributions}}
In summary, the main goal of this paper is to develop and experimentally validate an MPC-based real-time control scheme that leverages flexibility from EVCS and other controllable resources to track a pre-defined power profile at the grid connection point (GCP) of an ADN. The pre-defined power profile is computed by day-ahead formulation discussed in Part-I paper. The key contributions are listed below.
\begin{enumerate}
    \item We develop a real-time MPC for tracking the dispatch plan while accounting for the flexibility of EVCS and BESS and the uncertainty of the uncontrollable demand and generation. Compared to the MPC formulations in \cite{gupta2020grid, gupta2022reliable}, the proposed scheme models and accounts for the flexibility of EVCSs.
    \item We experimentally validate the proposed RT-MPC on a real-life ADN hosting multiple controllable units (two EVCS and BESS) and uncontrollable demand and PV plants. We demonstrate the {reliability and} continuity of the proposed MPC scheme by carrying out the experiment for multiple {consecutive} days.
\end{enumerate}

The structure of the paper is as follows. In Section~\ref{sec:Prob_stat}, we report the problem statement. Section~\ref{sec:methods} describes the methods for the real-time MPC scheme. Section~\ref{sec:expt_vald} reports the experimental validation of the proposed MPC scheme and finally, Section~\ref{sec:conclusion} concludes the work. 
%%%% 
\section{Problem Statement}
\label{sec:Prob_stat}
\begin{figure}[!htbp]
    \centering
    \includegraphics[width=0.9\linewidth]{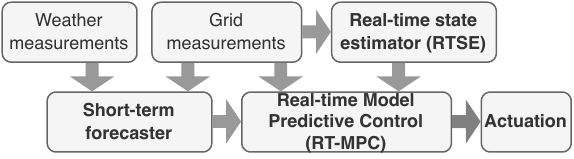}
    \caption{Overview of different processes during the real-time operation. On the top left, the measurement system (providing weather and grid measurements) feeds the RTSE, the short-term forecaster, and the RT-MPC layer. The RT-MPC layer sends power setpoints to the actuation layer.}
    \label{fig:overview}
\end{figure}

We consider an ADN with a generic topology (meshed or radial) {hosting} several EVCSs and other heterogeneous distributed energy resources (DERs) such as BESS, uncontrollable photovoltaic (PV) plants, and demand. The objective is to track a pre-determined power schedule (the dispatch plan) at the GCP of the ADN using the flexibility from the EVCSs and BESS. We propose a real-time model predictive control (RT-MPC)-based approach, which accounts for the short-term forecast of the uncontrollable injections and models of the controllable resources and tracks the dispatch plan accurately at a user-defined timescale.
The proposed scheme is experimentally validated on a real-life ADN interfaced with two EVCSs, two BESSs, loads, and multiple distributed PV generation plants. The RT-MPC consists of several building blocks. They are schematically shown in Fig.~\ref{fig:overview} and described below.
%%%%%%%%%
\begin{itemize}
    \item \textbf{Short-term forecasting:} provides predictions of the active and reactive power injections of the stochastic resources that are used in the real-time controller. These forecasts are updated {every second for the MPC horizon}. The scheme is described in Sec.~\ref{sec:short_forecast}.
    \item \textbf{Real-time state estimator (RTSE):} provides the most recent state of the grid (i.e., the nodal voltage and lines current phasors). These inputs are used in the RT-MPC and also in short-term forecasts. The RTSE is fed by a network of PMUs providing highly sampled voltage and line current measurements. The scheme is described in Sec.~\ref{sec:S_awareness}.
    \item \textbf{Real-time model predictive control:} computes the optimal active and reactive power setpoints for the controllable resources (i.e., the EVCS and BESS). It solves a constrained optimization problem with the objective to track the day-ahead dispatch plan while satisfying the constraints of the grid and {those of} the controllable resources. It is described in Sec.~\ref{sec:RTMPC}.
    \item \textbf{Actuation:} it receives the computed setpoints from the RT-MPC layer, verifies its feasibility based on the nominal ratings of the converter, and sends it to the controllable resources for the actuation. % It is described in Sec.~\ref{}.
\end{itemize}

The link between each block is schematically shown in Fig.~\ref{fig:overview}. The grid and the weather measurements feed the RTSE, the forecaster, and the RT-MPC. Finally, RT-MPC sends the power setpoints to the actuation layer.

%%%%%%%%
\section{Methods for Real-time Operation}
\label{sec:methods}
The two important components of the real-time operation are the \emph{short-term forecasting} {of the stochastic resources} and the  \emph{real-time MPC problem}. Each is described in the following subsections. 
\subsection{Short-term forecasting}
\label{sec:short_forecast}
The short-term uncertainty of the uncontrollable power injections such as the demand and PV generation is modeled by {dedicated} forecasts. Although the forecasting methods described are not the focus of the presented work, we describe them here for the sake of completeness. 
\subsubsection{Uncontrollable Demand}
To generate demand forecasts, we rely on a two-step scheme. First, we use the day-ahead forecasts (as described in Sec.~III-A-2, Part-I paper) for the current time step. The day-ahead forecast is up-sampled by linear interpolation from 5 minutes (day-ahead time sampling) to 30 seconds (real-time time sampling).
In the second step, we use a {persistent}\footnote{{The persistent forecast policy relies on real-time measurements available from PMUs. As these measurements are available almost in real-time, such a forecasting policy is capable to predict the immediate future demand with good accuracy.}} forecasting strategy, in which the first timestep of the forecast is replaced by the recent power measurements (using the grid states from RTSE). {The active and reactive power forecasts are denoted by $\mathbf{\widehat{p}}^\text{load}_{t}$ and $\mathbf{\widehat{q}}^\text{load}_{t}$, respectively.}

%%%%
\subsubsection{PV Generation}
To generate the PV forecasts, we rely on the measurements of the Global Horizontal Irradiance (GHI) and air temperature {integrated} with the {knowledge of the PV plants configuration}. It uses a similar PV model \cite{sossan2019solar} as described in Sec.~III-A-3 {of the} Part-I paper. 
{Furthermore}, we use persistent forecasting (i.e., replacing the first timestep forecast by the measurement itself). {The active and reactive power forecasts obtained are denoted by $\mathbf{\widehat{p}}^\text{pv}_{t}$ and $\mathbf{\widehat{q}}^\text{pv}_{t}$, respectively.}
\subsubsection{EV} forecast relies on the data input by EV consumers and the data sent by EVCS software \cite{EvTec_2}. For each charging session, the EV user provides the state-of-charge (SoC) at the arrival time, the expected stay duration, the preferred SoC target, and the capacity of the EV battery in power and energy. The EVCS software sends peak power capacity for each time step.
For the MPC horizon, we use persistent forecasting {with fast refresh-rate} measurements and update the forecasts at each time step. {The EV forecast consists of predictions on arrival time and the SoC target is defined later.}
%%%

In the following, we describe how these forecasts are used within the RT-MPC problem.
%%%%%%%%%%%%%%%
%%%%%%%%%%%%%%
\subsection{Real-time Model Predictive Control}
\label{sec:RTMPC}
\begin{figure}[!htbp]
    \centering
    \includegraphics[width=\linewidth]{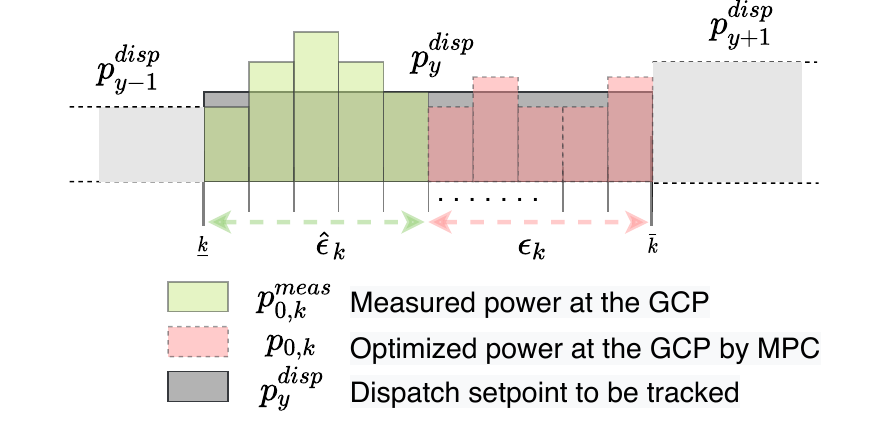}
    \caption{Schematic {representation} of the measured quantities and optimized variables in the RT-MPC. The implemented power at the GCP shown in light green color and the dispatch setpoint in grey color. The anticipated power at the GCP with RT-MPC is shown in red.}
    \label{fig:dispatch_error}
\end{figure}

%%%%%%%%%%%
The objective of the RT-MPC is to {control the} power setpoints {of} the controllable resources {to track} the day-ahead dispatch plan every 5 minutes for the whole day. 
Before, describing the problem formulation, we define the following notations.
\begin{itemize}[leftmargin=*]
    \item The dispatch setpoint to track is sampled at {5-minute} resolution\footnote{{It is consistent with day-ahead electricity market time resolution.}} and is denoted by $p^{disp}_y$, where the 5-minutes indices are denoted by \textcolor{black}{$y =  1, \dots, N-1$}, $N=288$ for 24 hours in a day. 
    \item The time resolution of the real-time control is 30-seconds; the index for real-time control is denoted by $k = 1, \dots, 2880$ (i.e., 10 time indices per 5-minutes dispatch period). 
    The first and the last 30-second\footnote{{It is decided based on the time taken to solve RT-MPC problem and communication overheads.}} indices during the 5-minute interval are denoted, respectively, by $\underline{k}$ and $\bar{k}$ (i.e., $\underline{k} = {\lfloor{\frac{k}{10}}\rfloor}\times10$ and $\bar{k} = \underline{k} + 10 - 1$). %Intra-
    \item The dispatch setpoint to be tracked at any time index $k$ is %can be computed by linking the day-ahead time resolution of 5~minutes with the real-time index of 30~seconds as 
    \begin{align}
        \bar{p}^{disp}_k = p^{disp}_{\lfloor{\frac{k}{10}}\rfloor},
    \end{align}
   where $\lfloor{.}\rfloor$ refers to the floor function. 
    \item The power measurements at the GCP, denoted by $p_{0,k}^{meas}$, are obtained and 
   the dispatch-energy error at time $k$ is computed by summing up the \textit{(i)} uncovered-energy errors from time index $\underline{k}$ to $k-1$,
    % \sum_{j = \underline{k}}^{k-1} (\bar{P}^{disp}_{j}-{P}^{meas}_{0,j})$
\begin{align}
   & \hat{\epsilon}_{k} = %\frac{1}{k-\underline{k}}
   \frac{30 (k+1-\underline{k})}{3600}\sum_{j = \underline{k}}^{k-1} (\bar{p}^{disp}_{j}-{p}^{meas}_{0,j})
   \label{eq:tracking_error_meas}
\end{align}
and 
\textit{(ii)} the anticipated error from $k$ to $\bar{k}$ given as
\begin{align}
&\epsilon_k = \frac{30(\bar{k}-k)}{3600}\sum_{j=k}^{\bar{k}}(\bar{p}^{disp}_{j}-{p}_{0,j}).
\label{eq:tracking_error_future}
\end{align}
Both energy-related errors $\epsilon_k$ and $\hat{\epsilon}_{k}$ are schematically shown in Fig.~\ref{fig:dispatch_error}. 
The implemented power at the GCP shown in light green color differs from the dispatch setpoint (shown in grey color) due to the inherent nature of uncertainty on the demand and generation, which is quantified by $\hat{\epsilon}_k$ in \eqref{eq:tracking_error_meas} to be compensated in {the} left-over MPC horizon period. A schematic representation of the timeahead setpoints computed by the MPC (until the end of the MPC horizon) is shown in red and expressed using decision variables by $\epsilon_k$ in \eqref{eq:tracking_error_future}. MPC optimizes EVCS and BESS setpoints and indirectly optimizes GCP power $p_{0,j}$ such that the sum $\epsilon_k + \hat{\epsilon}_{k}$ {tend to reduce} to zero by the end of the MPC horizon period.
\end{itemize}

The grid objective at a time index $k$ can be defined as minimizing the energy error incurred over a 5-minute horizon length with power setpoints actuated every 30~seconds (i.e, the dispatch energy error incurred at the GCP from current timestep to the end of the 5 min period).
%%%%%
\begin{align}
f^{disp}_k =  \epsilon_k + \hat{\epsilon}_{k}
\end{align}

The MPC problem considers the constraints of the grid and that of the controllable resources. In the following subsections, we describe the grid constraints and controllable resource models.
%%%%%%%%%%%%
\subsubsection{Grid constraints}
{Similar to the day-ahead stage (Part-I),} the grid constraints modeled by a linearized {power flow} grid model. They are {given below.}
\begin{subequations}
\label{eq:grid_cons}
\begin{align}
& \underline{\mathbf{v}} \leq  \mathbf{A}_{k}^\mathbf{v} \begin{bmatrix}
    \mathbf{p}_{k}\\
    \mathbf{q}_{k}
\end{bmatrix} + 
\mathbf{b}_{k}^\mathbf{v} \leq \bar{\mathbf{v}}
\label{eq:lin_grid_model_v}
\end{align}
\begin{align}
& \mathbf{0} \leq  \mathbf{A}_{k}^{\mathbf{i}} \begin{bmatrix}
    \mathbf{p}_{k}\\
    \mathbf{q}_{k}
\end{bmatrix} + 
\mathbf{b}_{k}^\mathbf{i} \leq \bar{\mathbf{i}}
\label{eq:lin_grid_model_i}
\end{align}
\begin{align}
    \begin{bmatrix}
    p_{0,k}\\
    q_{0,k}
\end{bmatrix}=\mathbf{A}_{k}^0 \begin{bmatrix}
    \mathbf{p}_{k}\\
    \mathbf{q}_{k}
\end{bmatrix} + \mathbf{b}_k^0; ~  (p_{0,k})^2  + (q_{0,k})^2 \leq S_{\text{max}}^2.
\label{eq:lin_grid_model_pl}
\end{align}
Eq. \eqref{eq:lin_grid_model_v} constrains the nodal voltages ($\mathbf{v} \in \mathds{R}^{(n_b-1)}$, $n_b$ being number of nodes in the grid), by bounding them to upper and lower limits $[\underline{\mathbf{v}}~\bar{\mathbf{v}}]$. Eq.~\eqref{eq:lin_grid_model_i} expresses the lines current magnitudes ($\mathbf{i} \in \mathds{R}^{n_l}$, $n_l$ being the number of lines) and bounds them to their respective ampacities $\bar{\mathbf{i}}$. Eq.~\ref{eq:lin_grid_model_pl} expresses the GCP power and bounds it to the transformer capacity $S_{\text{max}}$.
Note that $\mathbf{p} \in \mathds{R}^{(n_b-1) }$ and $\mathbf{q} \in \mathds{R}^{(n_b-1) }$ are the three-phase total nodal active and reactive controllable injections for all nodes except the slack node. $\mathbf{A}^\mathbf{v} \in \mathds{R}^{(n_b-1)\times 2(n_b-1)}$ and $\mathbf{b}^\mathbf{v} \in \mathds{R}^{(n_b-1)}$, $\mathbf{A}^{\mathbf{i}} \in \mathds{R}^{n_l\times 2(n_b-1)}$ and $\mathbf{b}^{\mathbf{i}}  \in \mathds{R}^{n_l}$ are the coefficients derived from the voltage magnitude sensitivity coefficients and operating point (as described in \cite{GuptaThese}), $\mathbf{A}^{0} \in \mathds{R}^{2 \times 2(n_b-1)}$ and $\mathbf{b}^{0} \in \mathds{R}^2$ being the coefficients corresponding to the GCP power.
The power injections in \eqref{eq:grid_cons} consist of controllable and uncontrollable powers at the nodes for time index $k$ {containing the contributions of all the resources. It is worth noting that it may contain several resources per node as well. It is} given as
%\begin{subequations}
\begin{align}
    & \mathbf{p}_{k} = \mathbf{\widehat{p}}^\text{load}_{k} - \mathbf{\widehat{p}}^\text{pv}_{k} +  \mathbf{p}^\text{evcs}_{k} + \mathbf{p}^\text{bess}_{k} \\
    & \mathbf{q}_{k} = \mathbf{\widehat{q}}^\text{load}_{k} - \mathbf{\widehat{q}}^\text{pv}_{k} +  \mathbf{q}^\text{evcs}_{k} + \mathbf{q}^\text{bess}_{k}
\end{align} 
\end{subequations}
where {$\mathbf{\widehat{p}}^\text{load}_{k}/\mathbf{\widehat{q}}^\text{load}_{k}$, $\mathbf{\widehat{p}}^\text{pv}_{k}/\mathbf{\widehat{q}}^\text{pv}_{k}$, $\mathbf{p}^\text{bess}_{k}/\mathbf{q}^\text{bess}_{k}$ and $\mathbf{p}^\text{evcs}_{k}/\mathbf{q}^\text{evcs}_{k}$ are the nodal active/reactive} power injections corresponding to load, PV generation, BESS and EVCS, respectively.

\subsubsection{EVCS objectives and constraints}
Let the EVCS be located at indices $i \in \mathcal{N}^{ev}$ where $\mathcal{N}^{ev}$ denotes the set of node indices where EVCSs are installed. Let each EVCS has multiple {charging} plugs indexed by $l = 1, \dots, L_i$. 
Similar to the day-ahead formulation, the EVCS cost function has multiple objectives: the first term minimizes the worst-case difference between the SoC target and SoC that can be achieved at the end of the MPC horizon, the second term minimizes the EV battery wearing between subsequent time-steps \cite{yao2016real} (i.e., $\left| p_{k+1,l,i}^{\text{evcs}} - p_{k,l,i}^{\text{evcs}} \right|$, where $p_{k,l,i}^{\text{evcs}}$ is the power demand at indices $k, l, i$).
The EVCS objective at time index $k$ is given by
\begin{align}
\label{eq:EV_obj}
\begin{aligned}
   f^{evcs}_k =  & \frac{3600}{(\delta{t}) L_i} \sum_{i\in\mathcal{N}^{ev}}\sum\limits_{l=1}^{L_i} {\max \left\lbrace {\Big(\text{SoC}^{\text{Target}}_{l,i,\bar{k}}}- \text{SoC}^{\text{evcs}}_{l,i,\bar{k}} \Big), 0 \right\rbrace} + \\
    & \frac{1}{(\bar{k} - k)L_i}\sum_{i\in\mathcal{N}^{ev}} \sum\limits_{l=1}^{L_i} \sum\limits_{k=k+1}^{\bar{k}} \left| p_{k+1,l,i}^{\text{evcs}} - p_{k,l,i}^{\text{evcs}} \right|
\end{aligned}
\end{align}
where $\delta t$ is the sampling time (i.e., 30~sec). 

Since the RT-MPC is solved with short MPC horizon of 5~minutes, the SoC target needs to be {updated} for the end of 5~minutes. 
This is done by interpolation using the current SoC, target SoC, and departure time. SoC target time index $k$ is given by
\begin{align}
    \text{SoC}^{\text{Target}}_{l,i,\bar{k}} = \text{SoC}^{\text{evcs, meas}}_{l,i,k-1} + (\text{SoC}^{\text{Target}}_{l,i} - \text{SoC}^{\text{evcs, meas}}_{l,i,k-1}) \frac{\bar{k} - k}{k_f - k}
\end{align}
where $\text{SoC}^{\text{evcs, meas}}_{l,i,k-1}$ is the measured SoC at time index $k-1$, $\text{SoC}^{\text{Target}}_{l,i,\bar{k}}$ the target SoC for time index $k$, $l-$th plug, $i-$th EVCS and the departure time index $k_f$.

The EVCS constraints consist in limiting on the EV state-of-charge ($\text{SoC}^{\text{evcs}}_{k}$) and are expressed as
\begin{subequations}
\label{eq:evcs_cons}
% \begin{align}
%     & {0 \leq \text{SoC}_{t,k,i,\omega} \leq 1} 
% \end{align}
% where, 
\begin{align}
    & 0 \leq  \text{SoC}^{\text{evcs}}_{k,l,i} = \text{SoC}^{\text{evcs}}_{k-1,l,i} - \frac{p_{k,l,i}^\text{evcs} (\delta t)}{E_{l,i}^{\text{max}}} \leq 1 
\end{align}
where ${E_{l,i}^{\text{max}}}$ is the energy capacity for EV connected to plug $l$ and node $i$.
Another constraint limits the EVCS active power by the charger's minimum and maximum admissible\footnote{The minimum and maximum admissible setpoint per plug is sent by the charger to the MPC controller.} setpoint given by $p^{\text{evcs},\text{min}}_{k,l,i}$ and $p^{\text{evcs},\text{max}}_{k,l,i}$, respectively.
\begin{align}
        & \mu_{k,l,i}~p^{\text{evcs},\text{min}}_{k,l,i} \leq p_{k,l,i}^\text{evcs} \leq \mu_{k,l,i}~p^{\text{evcs},\text{max}}_{k,l,i} 
\end{align}
\end{subequations}
where $\mu_{k,l,i}$ is a \emph{known} boolean expressing whether, or not, an EV is connected to plug $l = 1,...,L_i$. The reactive power from the EVCS is uncontrollable and null (i.e., $q_{k,l,i}^\text{evcs} = 0, \forall k, l, i$).
%%%%%%%
%%%%%%%%%
%%%%%%%%%%%%%%%%%%%%
\subsubsection{BESS Objectives and Constraints}
The BESS is one of the controllable resources in the proposed dispatching framework. It can provide both active and reactive power regulation during real-time operation. Let $p^{\text{bess}}_{k,i}, q^{\text{bess}}_{k,i}$ be the battery's active and reactive power at time $k$ and node $i\in\mathcal{N}^{bess}$ where $\mathcal{N}^{bess}$ is the set of node indices where BESSs are installed. 

For each BESS, the objective is to minimize its usage (i.e., absolute injections $|p^{\text{bess}}_{k,i}|$), to mitigate its cycling related aging. The objective is 
\begin{align}\label{eq:bess_obj}
   f^{bess}_k =  \sum_{i\in\mathcal{N}^{bess}}\sum_{k = k}^{\bar{k}}|p^{\text{bess}}_{k,i}|
\end{align}

It considers the constraints on the state-of-charge i.e., $\text{SoC}^{\text{bess}}_{k,i}$ of each battery ($i\in\mathcal{N}^{bess}$) is constrained by SoC limits [$\underline{\text{SoC}}^{\text{bess}}_i,~\overline{\text{SoC}}^{\text{bess}}_i$],  ${\underline{\text{SoC}}^{\text{bess}}_i}/{\overline{\text{SoC}}^{\text{bess}}_i}$ are the minimum/maximum SoC bounds of battery $i$
\begin{subequations}
\label{eq:bess_cons}
\begin{align}
    {\underline{\text{SoC}}^{\text{bess}}_i} \leq \text{SoC}^{\text{bess}}_{k,i} = \text{SoC}^{\text{bess}}_{k-1,i} - \frac{p^{\text{bess}}_{k,i} (\delta t)}{E^{\text{bess, max}}_i} \leq {\overline{\text{SoC}}^{\text{bess}}_i}. 
    \label{eq:SOE_update_dis}
\end{align}
where $E^{\text{bess, max}}_i$ represent EV energy capacity.
Moreover, the active and reactive power are bounded by converter ratings
\begin{align}
& (p^{\text{bess}}_{k,i})^2 + (q^{\text{bess}}_{k,i})^2 \leq ({S}^{\text{bess,max}}_i)^2  \label{eq:BETT_cap_dis}
\end{align}
\end{subequations}
where ${S}^{\text{bess,max}}_i$ is the rated power of the converter associated to the i-th battery. The circle constraint in \eqref{eq:BETT_cap_dis} is approximated by a set of piece-wise linear constraints as in \cite{nick2014optimal}.

\subsubsection{Final MPC Formulation}
The final MPC problem at time index $k$ is formulated as
\begin{subequations}
\label{eq:RT_mpc_form}
 \begin{align}
    & \underset{\mathbf{p}_k^{\text{evcs}}, \mathbf{p}_k^{\text{bess}}}{\text{minimize}}~f^{disp}_k + f^{evcs}_k + f^{bess}_k\\
    & \text{subject to}~~~~\eqref{eq:grid_cons}, \eqref{eq:evcs_cons}, \eqref{eq:bess_cons}.
\end{align}   
\end{subequations}

As it can be observed, the problem is linear, thanks to the linearized grid model and piece-wise linear approximation of the capability curves \eqref{eq:lin_grid_model_pl} and \eqref{eq:BETT_cap_dis}. Therefore, the problem can be solved efficiently by any LP solver. 

{It may appear that the RT-MPC formulation in \eqref{eq:RT_mpc_form} is similar to the day-ahead problem from the Part-I paper. However, the key difference are listed below: 
\begin{itemize}
    \item The RT-MPC formulation is designed to achieve real-time tracking of the dispatch schedule decided by the day-ahead stage. It is formulated as model predictive control to account for the real-time variations of the prosumption, whereas the day-ahead stage is formulated as scenario-based stochastic optimization scheme.
    \item The RT-MPC optimization problem is much faster to solve (below 30 seconds), whereas the day-ahead formulation can take a few hours to solve.
\end{itemize}}

\section{Experimental Validation}
\label{sec:expt_vald}
In this section, we present the experimental validation of the described RT-MPC on a real-life ADN. In particular, we first describe the hardware setup, the monitoring, and communication infrastructure, and then present the experimental validation results for different day types.
%%%%%%%%%%%%%%%%%%%%%%%%%%%%%%%
%%%%%%%%%%%%%%%%%%%%%%%%%%%%%%%
%%%%%%%%%%%%%%%%%%%%%%%%%%%%%%%
\subsection{{Experimental Infrastructure}}
\label{sec:sim_setup}
\begin{figure}[!htbp]
    \centering
    \includegraphics[width=\linewidth]{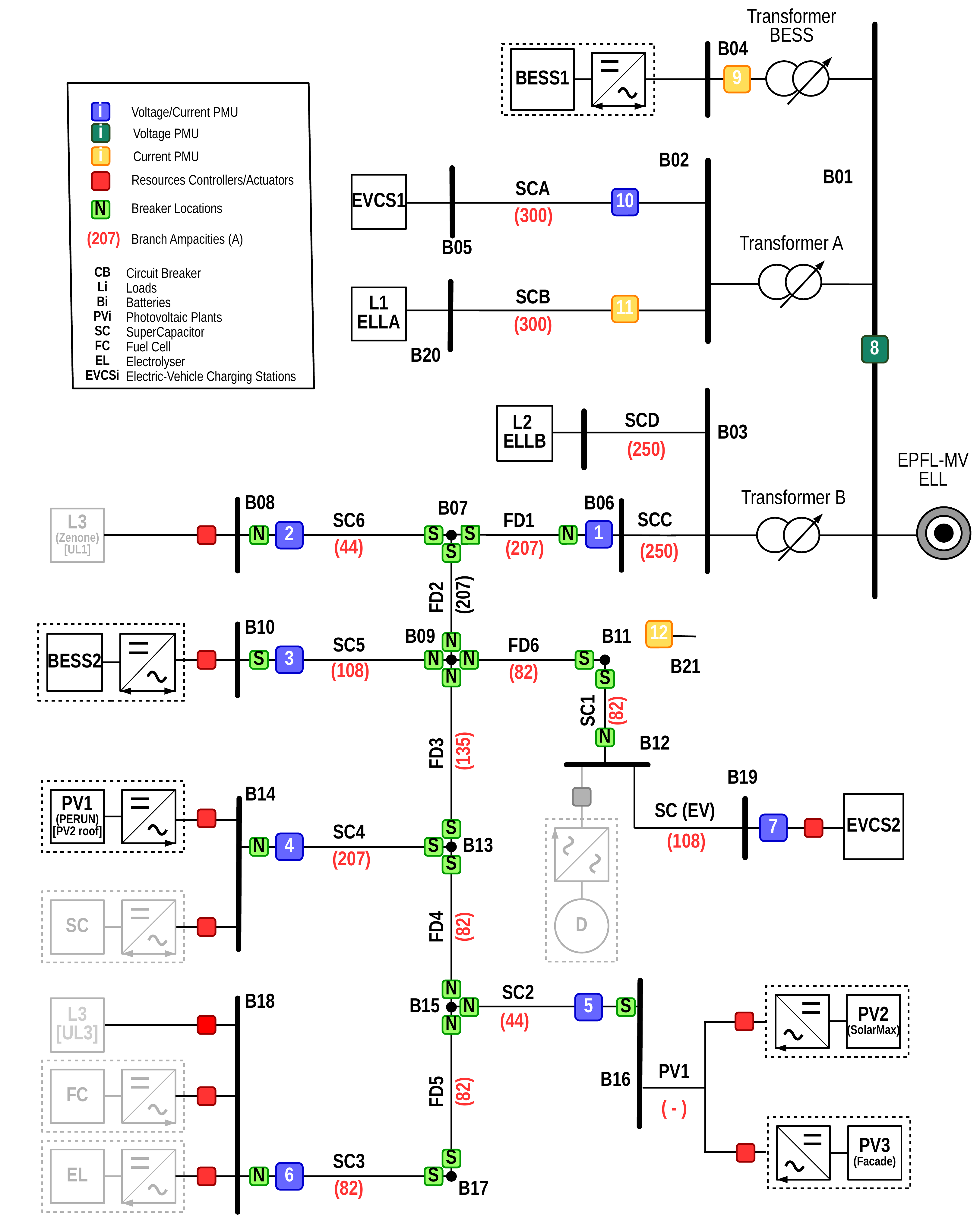}
    \caption{{Schematic representation of the experimental infrastructure of the} ELL building, EPFL.}
    \label{fig:ELL_network}
\end{figure}

The experimental validation is performed on the setup described in Part-I of the paper, shown in Fig.~\ref{fig:ELL_network}. The detailed description of the resources and the grid is included in Part-I of the paper. 

For the experimental validation of the RT-MPC, we consider two BESS and two EVCS as controllable resources. Other resources such as PV plants and buildings are treated as uncontrollable. Each resource used in the experiment is described below.

%%%%%
\begin{figure}[!htbp]
    \label{fig:BESSboth}
    \centering
    \subfloat[BESS~1]{\includegraphics[width=0.45\linewidth]{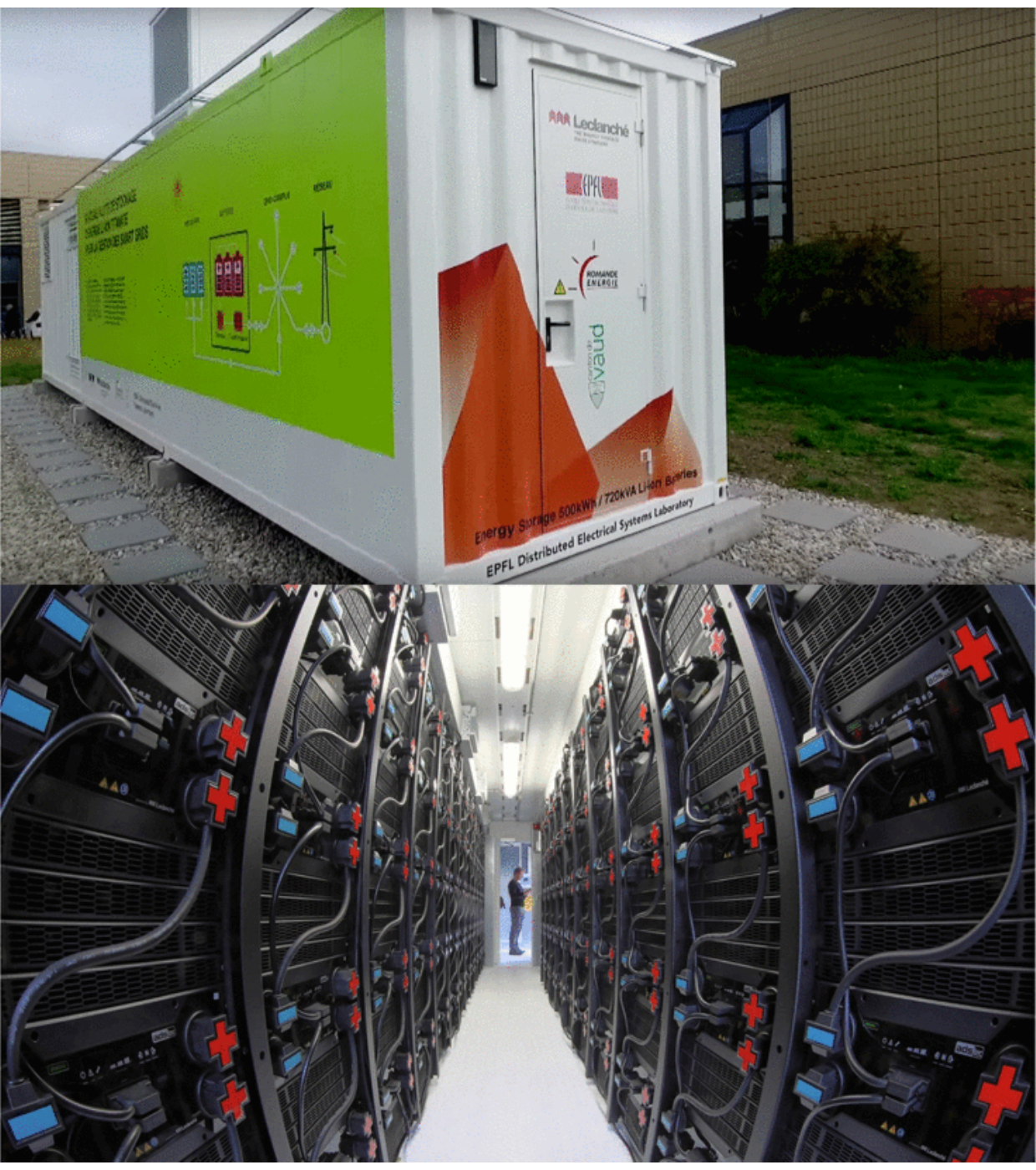}\label{fig:BESS2}} 
    \subfloat[BESS~2]{\includegraphics[width=0.38\linewidth]{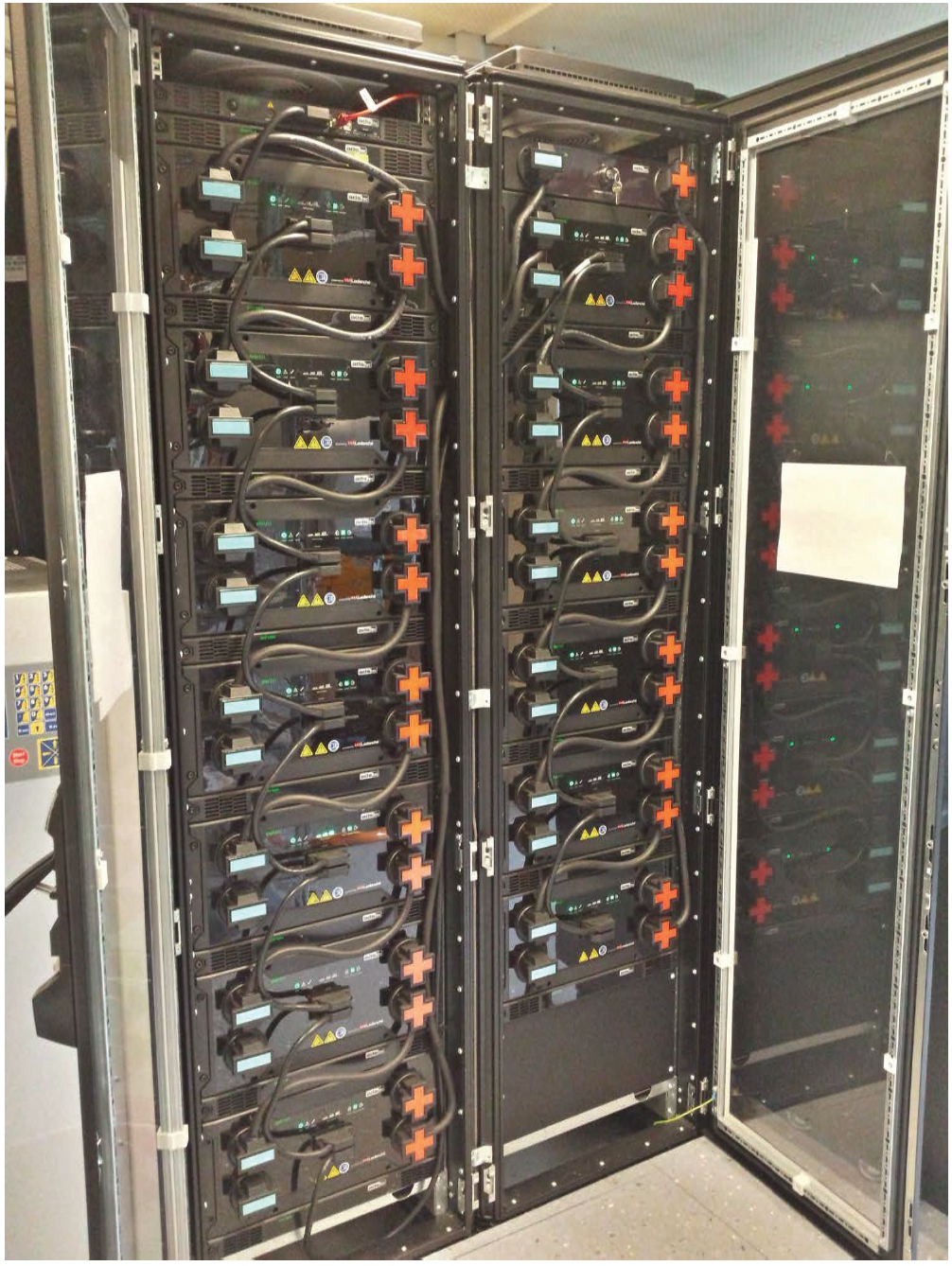}\label{fig:BESS1small}}
    \caption{Two battery storage installations used for the real-time experiments. BESS 1 and BESS 2 are connected to B04 and B10 respectively.}
\end{figure}
%%%%
\begin{figure}[!htbp]
\centering
\subfloat[EVCS1 (Gofast)]{\includegraphics[width=0.42\linewidth]{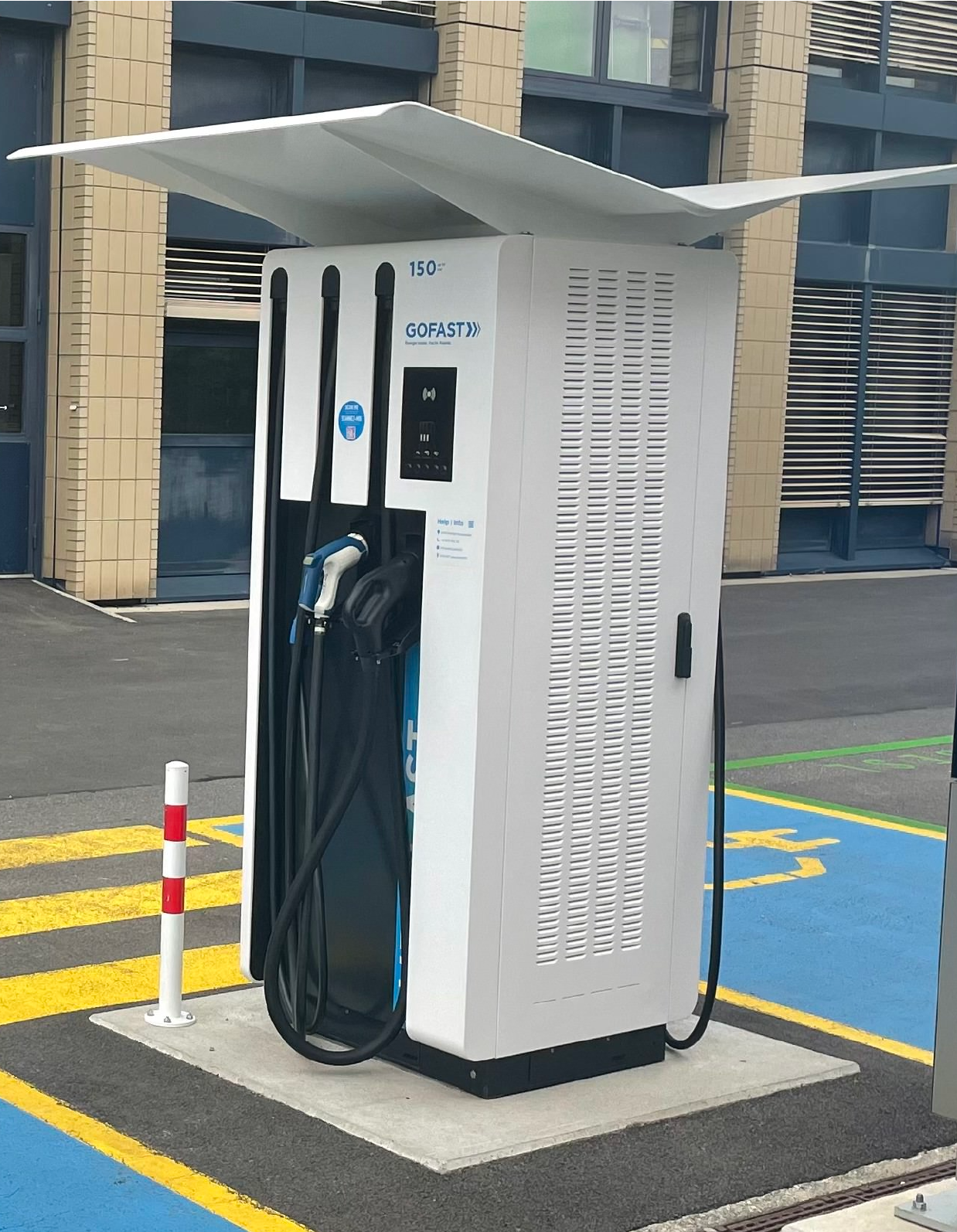}}\label{fig:EVCS1}
\subfloat[EVCS2]{\includegraphics[width=0.4\linewidth]{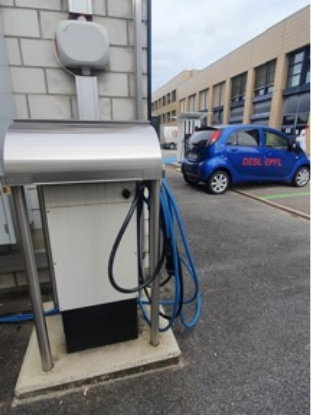}\label{fig:EVCS2}} 
\caption{(a) Level-3 Gofast EVCS with 5 plugs and (b) Level-2 EVCS with 3 plugs.}
\label{fig:EVCS_installation}
\end{figure}

%%%%
\begin{figure}[!htbp]
    \centering
   \includegraphics[width=0.6\linewidth]{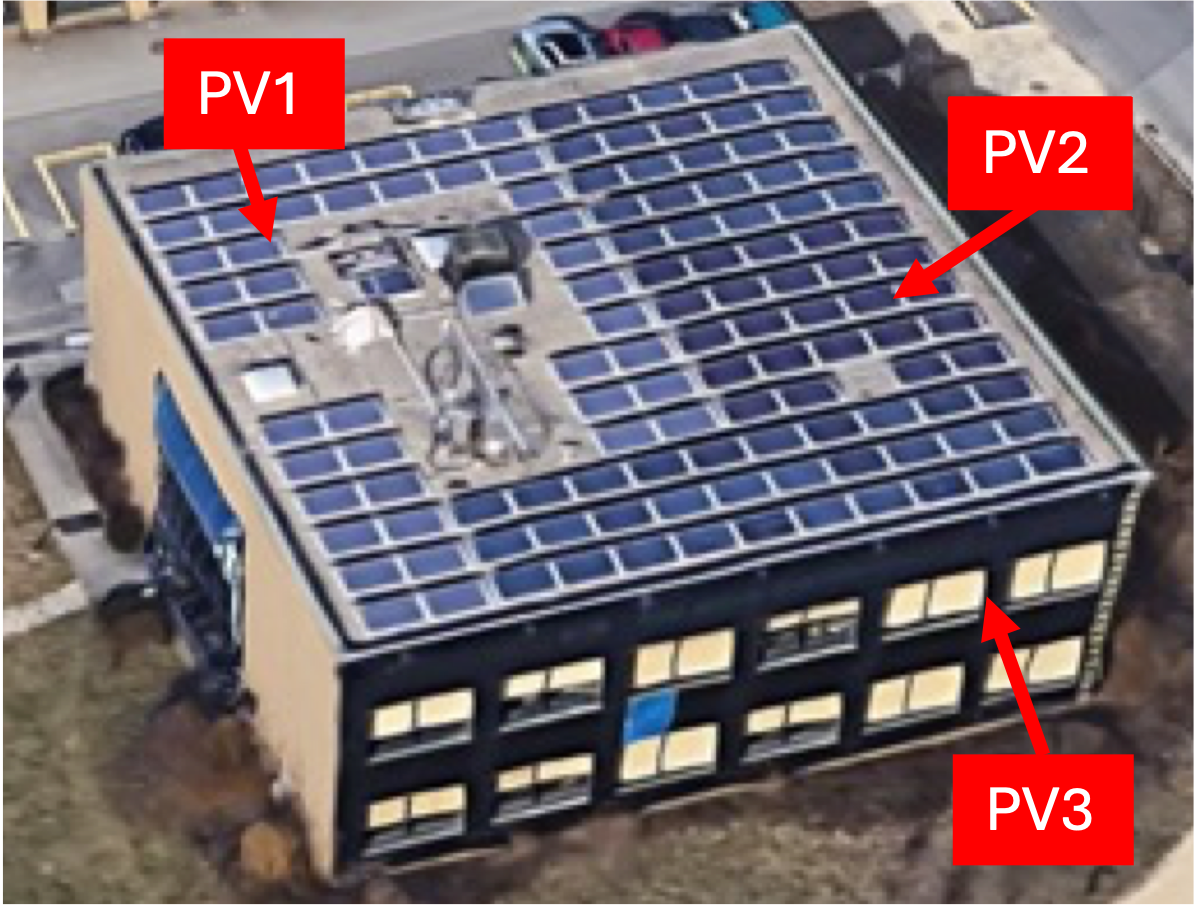}
    \caption{Rooftop PV plants (PV1 and PV2) and Facade PV plant (PV3).}
    \label{fig:PV_plants}
\end{figure}
%%%%%
\begin{itemize}
    \item \textbf{BESS}: the ADN hosts two different BESS with capacities of 25kW/25KWh (BESS1) and 150kW/300kWh (BESS2). Both BESS are based on the Lithium Titanate Oxide {(LTO)} and are rated for 15,000 cycles (of 1C/1C charge/discharge rates and 100~\% depth of discharge). BESS1 consists of 13 modules, and BESS2 consists of 9 parallel racks where each rack consists of 15 modules. Each module is composed {by} 20s3p cells. A view of BESS1 and BESS 2 is shown in Fig.~\ref{fig:BESS1small} and \ref{fig:BESS2}.
    \item \textbf{EVCS}: the ADN hosts two different charging stations EVCS1 and EVCS2 as shown in Fig.~\ref{fig:EVCS_installation}. EVCS1 is a {level}-3 {DC fast} charging station and has a peak power capacity of 172kW. It hosts 5 plugs but only 2 plugs can be operational at the same time. EVCS2 is a {level-2 AC} charger with a 32 kWp capacity. It hosts 3 plugs, but only 2 plugs can be operational at the same time. The capacity per plug is reported in Table~\ref{tab:EVCS}. EVCS plugs, depending on their type, use different communication protocols. For example, the CHAdeMO plugs use the DCMS protocol from EvTec \cite{EvTec_2}. Type-2 plugs use IEC-61852 \cite{ies61851}.
\begin{table}[!htbp]
    \centering
    \caption{EVCS plug ratings.}
    \begin{tabular}{|c|c|c|}
    \hline
     \bf Resources      &  \bf Labels    & \bf Ratings  \\
     \hline
     Electric Vehicle & EVCS1      &   1 $\times$ Type-2 Plug - 43~kWp  \\
     Charging Station &  (5~plugs) & 1 $\times$ Type-2 Plug - 22~kWp  \\
                      &  {Level-3}            & 2 $\times$ CCS DC Plug - 150~kWp  \\
                     &              & 1 $\times$ CHAdeMO Plug - 150~kWp \\
                      \hline
                      & EVCS2   &   2 $\times$ Type-2 Plug - 22~kWp  \\
                      & (3~plugs)       &  1 $\times$ CHAdeMO Plug - 10~kWp \\
                      & {Level-3} & \\
    \hline
    \end{tabular}
    \label{tab:EVCS}
\end{table}

    Furthermore, as explained in Sec. III-C of \cite{rudnik2021experimental, FahmyThesis}, power-to-current lookup tables are needed to enable explicit active power control of {level-2} plugs. Recall that {level-2} plugs are controlled through an analog pulsed signal that dictates to the EV the RMS value of the maximum per-phase current it can consume. As a result, power-to-current lookup tables were precomputed. 
    For more details on the characterization of the current-power lookup table, the reader can refer to \cite{FahmyThesis}.
    \item \textbf{PV}: The grid hosts 3 different PV plants, PV1 and PV2 are rooftop PV rated at 13kWp and 16kWp, {respectively. PV3} is on a building facade and is rated at \textcolor{black}{13.2~kWp}. All these {power} generation {assets} are considered uncontrollable in the experiment and are operated at their maximum power potential (i.e., at the MPPT). The three PV plants are shown in Fig.~\ref{fig:PV_plants}.
    \item \textbf{Demand:} it refers to the electricity demand corresponding to the building shown in Fig.~\ref{fig:PV_plants}. The consumption is split over two transformers: L1-ELLA (i.e., first load) and L2-ELLB (i.e., second load) {associated to} transformers A and B, respectively and rated at \textcolor{black}{20~kW and 5~kW, respectively.} 
\end{itemize}
%%%
\subsection{Situational Awareness Infrastructure}
\label{sec:S_awareness}
The experimental setup is equipped with a dedicated monitoring and communication infrastructure enabling real-time state estimation (RTSE), and information of the resources state such as BESS SoC, current setpoints, EVCS power etc. In the following, we briefly describe them.
%%%
\subsubsection{Communication Infrastructure, Centralized Server and Data-Logging}
\begin{figure}[!h]
    \centering
    \includegraphics[width = 0.95\linewidth]{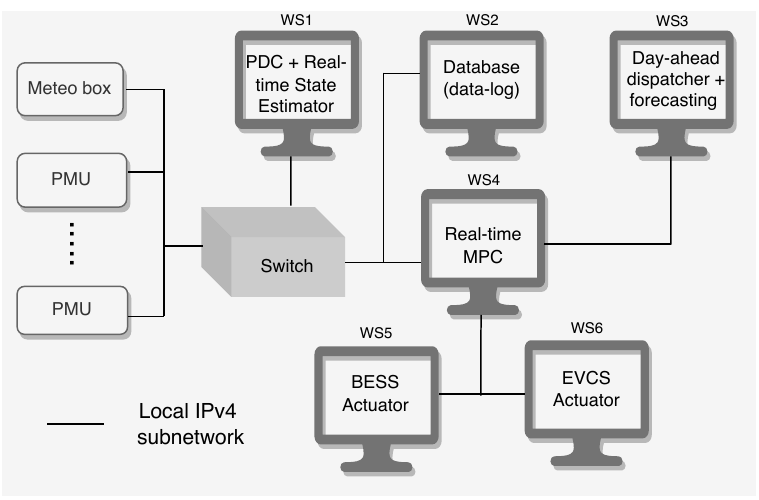}
    \caption{\textcolor{black}{IT communication infrastructure of the experimental setup.}}
    \label{fig:ITComm}
\end{figure}

ADN {sensing} and connected resources communicate over a dedicated IPv4 communication network. 
Fig.~\ref{fig:ITComm} shows the schematic of the communication network. It connects different {workstations (WS1-WS6)} hosting the RTSE, data server, day-ahead dispatcher, RT-MPC and the actuator. WS1 collects the PMU packets and {hosts} a phasor data concentrator (PDC) \cite{PMU_standard}; the data is then used {to estimate the state of the power grid in real-time each 20ms}. A discrete Kalman filter-based state estimator processes the measurements \cite{kettner2017sequential} and provides the estimates of the voltage and current phasors of all the nodes and lines with a total latency of less than 80~ms with respect to the UTC-GPS time tag of the PMU measurements. Details on the RTSE can be found in \cite{kettner2017sequential, zanni2020pmu}. WS2 implements a SQL-based (influxdb \footnote{https://www.influxdata.com}) database to log all the time-series during the experiments. WS3 computes the day-ahead forecasts and dispatch plan (formulation reported in Part-I paper). WS4 runs the real-time MPC, and computes short-term forecasts for different resources. WS5 and WS6 run the BESS and EVCS softwares, respectively to implement the power setpoints computed by the RT-MPC.
The workstations, the PMUs and meteobox are connected by a switch. 

\subsubsection{Time-tagged measurement infrastructure}
\begin{figure}[t]
\centering
\footnotesize
\subfloat[]{\includegraphics[width=0.35\linewidth]{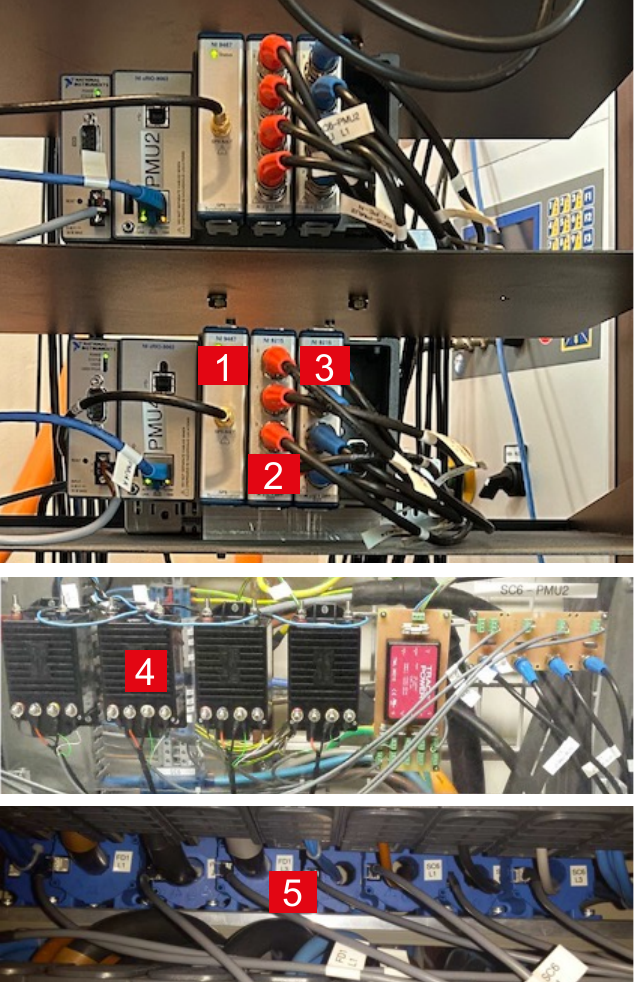}\label{fig:PMUs}}
\hspace{0.1em}
\subfloat[]{\includegraphics[width=0.35\linewidth]{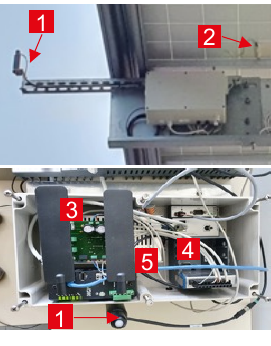}\label{fig:meteobox}} \\
\caption{(a) Installed PMUs implemented on National Instrument Compact RIO {hardware}, (1) GPS module, (2) voltage measurements acquisition module, (3) current measurement acquisition module (4) voltage transducers {LEM CV 3-1000} (5) current transducers {LF 205-S/SP3} and (b) GHI and temperature measurement box (Meteobox) at a PV plant: (1) pyranometer, (2) temperature sensor (3) {telecom} antenna (4) power supply (5) NI Compact RIO.} 
\end{figure}

As discussed before, the network is monitored by several PMUs that provide real-time information of the grid states (coupled to a RTSE) and a meteobox that provides real-time information on GHI and air temperature. These measurements are also stored in the data-server as historical measurements, which are used for day-ahead and short-term forecasting. The measurement units are briefly described below.
\begin{itemize}
    \item \textbf{PMU(s)}: are used to provide time-synchronised and time-tagged measurements at 50 frames per second. {An example} of the installed PMU is shown in Fig.~\ref{fig:PMUs}. The PMUs are implemented {on} National Instrument Compact RIO consisting of a GPS module and {current/voltage} acquisition modules. More details on the specification of the components are in \cite{ReyesThesis}
    \item \textbf{Meteobox}: is installed to get measurements of GHI, air- and PV-panel- temperatures. It provides real-time measurements with a sampling rate of 1~sec (including communication latency). Fig~\ref{fig:meteobox} shows the meteobox installed on the site; each meteobox consists of a pyranometer (to sense the GHI), two temperature sensors, and a power supply. More details on the specification can be found in \cite{EnricaThesis}.
\end{itemize}
%%%%
%%%%%%%%
\subsection{Dataflow}
Fig.~\ref{fig:Schematic_RTflow} shows the sequence of operations and communication flows during the real-time stage.
The real-time stage starts at 00:00 local time. First, we retrieve the current power dispatch setpoint $\bar{p}_k^{disp}$. Then, the grid measurements are obtained and used to {infer the actual system state (used to formulate the LOPF)}. It then updates the resources states (such of BESS and EV SoC). Then, the short-term forecasts of the uncontrollable resources are updated. All these information is then used for solving RT-MPC in \eqref{eq:RT_mpc_form}. The optimized power setpoints are then sent to EVCS and BESS actuators. This cycle is repeated every 30~sec until 23:59:30 local time.
%%%%

\tikzstyle{Block} = [rectangle, rounded corners, minimum width=3cm, minimum height=0.5cm,text centered, text width = 7cm, draw=black, fill=black!5]
\tikzstyle{Block1} = [rectangle, rounded corners, minimum width=1.5cm, minimum height=0.5cm,text centered, text width = 1.5cm, draw=black, fill=black!5]
\tikzstyle{Block2} = [rectangle, rounded corners, minimum width=5.0cm, minimum height=0.5cm,text centered, text width = 5.5cm, draw=black, fill=black!5]
\tikzstyle{arrow} = [thick, ->, >=stealth]

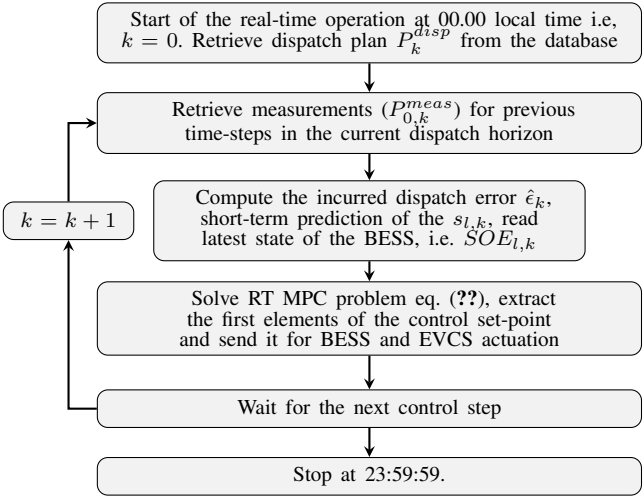
\begin{figure}
\centering
\begin{tikzpicture}[node distance=2cm]
    \node (a) [Block] {\vspace{-0.15cm}
    \footnotesize Start of the real-time operation at 00.00 local %\vspace{-0.15cm} 
    time i.e, $k=0$. Retrieve dispatch plan $P^{disp}_k$ from the database };
     \node (j) [Block, below of = a, yshift = 0.8cm] {\vspace{-0.1cm}
    \footnotesize \footnotesize Retrieve measurements ($P_{0,k}^{meas}$) for previous time-steps in the current dispatch horizon};
    %\node (b) [Block, below of = j, yshift = 0.75cm] {\vspace{-0.15cm} \footnotesize If start of the 5-minute dispatch horizon, update intra-day 5-minutes forecast \vspace{-0.15cm} using {Algorithm~\ref{alg:IntradayForecast}}, compute energy budget \vspace{-0.15cm} $\Delta SoE$ by solving upper layer MPC problem eq.~\eqref{eq:UpperMPC}};
    \node (c) [Block2, below of = j, yshift = 0.7cm] {\vspace{-0.15cm} \footnotesize Compute the incurred dispatch error $\hat{\epsilon}_k$, \vspace{-0.15cm} short-term prediction of the $s_{l,k}$, read latest state of the BESS, i.e. $SOE_{l,k}$};
    %\node (d) [Block2, below of = c, yshift = 1cm] {\vspace{-0.15cm} \footnotesize Compute the short-term prediction of the powers at all the nodes $s_{lt}$ by using RTSE};
   % \node (e) [Block2, below of = d, yshift = 1cm] {\footnotesize Read latest state of the BESS, i.e. $SoE_{l,k}$};
    \node (f) [Block, below of = c, yshift = 0.7cm] {\vspace{-0.15cm} \footnotesize Solve RT MPC problem eq.~\eqref{eq:lowerMPC}, %\vspace{-0.15cm} 
    extract \vspace{-0.15cm} the first elements of the control set-point and send it for BESS and EVCS actuation};
    \node (g) [Block, below of = f, yshift = 0.8cm] {\footnotesize Wait for the next control step};
    \node (h) [Block1, left of = c, xshift = -2cm] {\footnotesize $k=k+1$};
    \node (i) [Block, below of = g, yshift = 1.1cm] {\footnotesize Stop at 23:59:59.};
    \draw [arrow] (a) -- (j);
    \draw [arrow] (j) -- (c);
    %\draw [arrow] (b) -- (c);
    %\draw [arrow] (c) -- (d);
    %\draw [arrow] (d) -- (e);
    \draw [arrow] (c) -- (f);
    \draw [arrow] (f) -- (g);
    \draw [arrow] (g) -| (h);
    \draw [arrow] (h) |- (j);
    \draw [arrow] (g) -- (i);
\end{tikzpicture}
    \caption{Flow-chart showing real-time operation during 24 hours.}
    \label{fig:Schematic_RTflow}
\end{figure}

%%%%%%%%%

%%%%%%%%%%%%
%%%%%%%%%%%
\begin{figure}[!htbp]
    \centering
     \subfloat[Dispatch plan, and power at the GCP with and without control.]{ \includegraphics[width=1\linewidth]{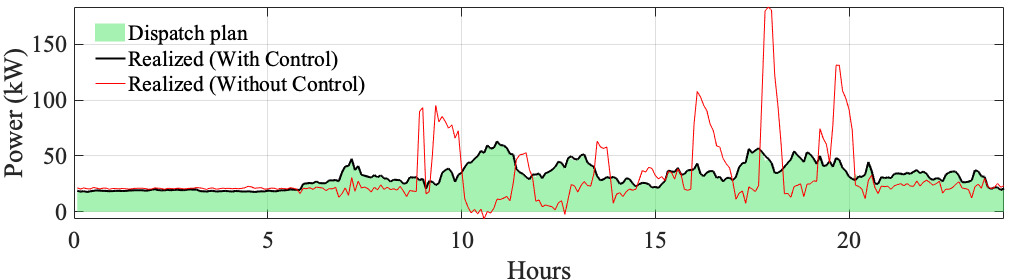}\label{fig:GCP_dispatch_day1}} \\
     \subfloat[Upper {panel}: active power regulation from the BESS1, and lower panel: {measured state-of-charge (SoC}).]{\includegraphics[width=\linewidth]{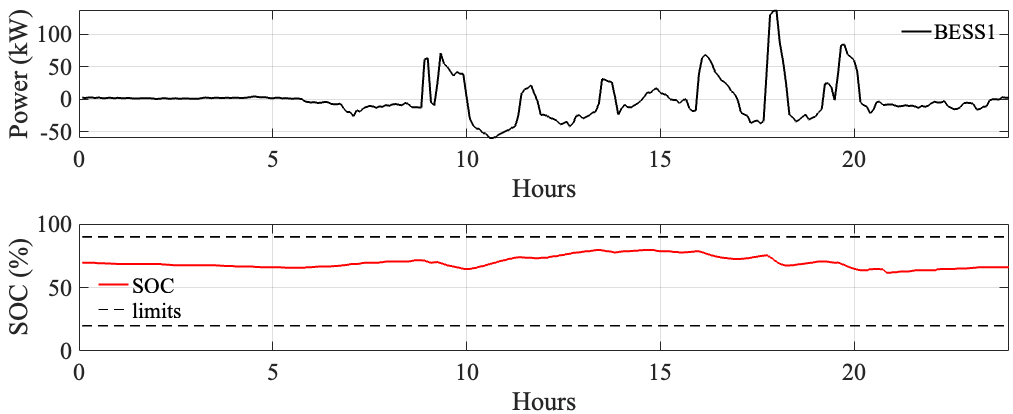}\label{fig:BESS_day1}}\\
     \subfloat[Upper {panel}: Controlled active power consumption by EVCS1, lower {panel}: EV SoC during the day along with the SoC target.]{\includegraphics[width=\linewidth]{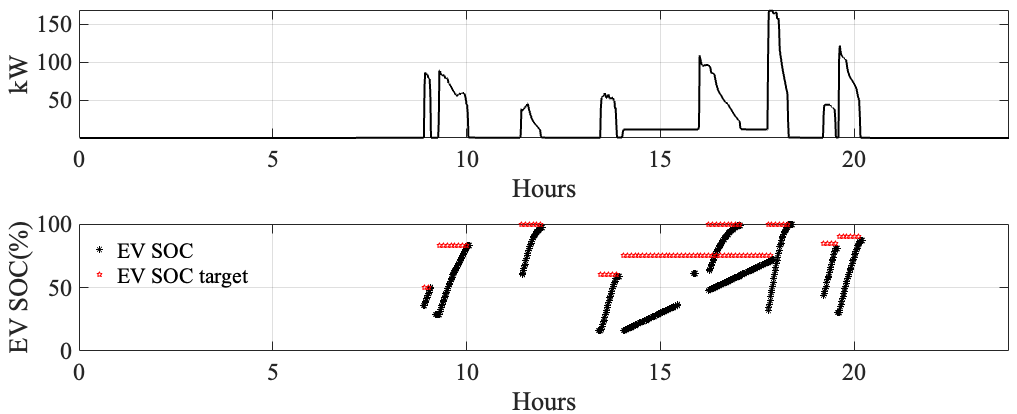}\label{fig:EVCS1_day1}}\\
    \subfloat[Upper {panel}: Controlled active power consumption by EVCS2, lower {panel}: EV SoC during the day along with the SoC target.]{\includegraphics[width=\linewidth]{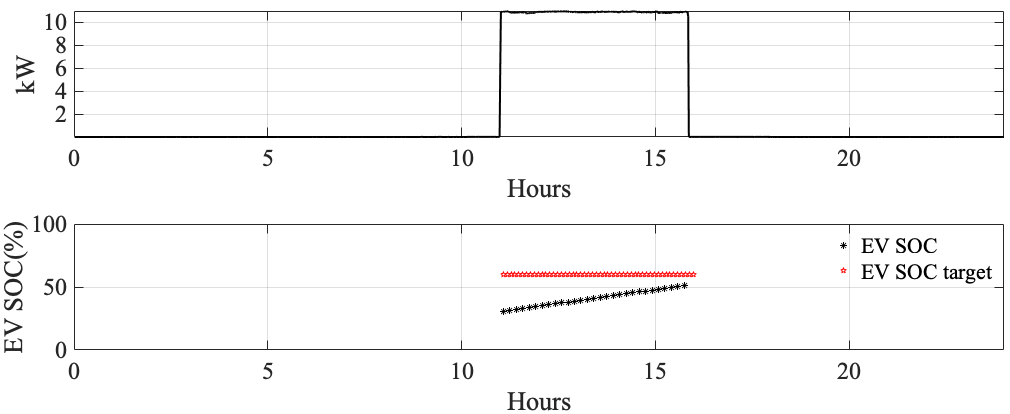}\label{fig:EVCS2_day1}}\\
    \subfloat[Measured demand at ELLA and ELLB.]{\includegraphics[width=\linewidth]{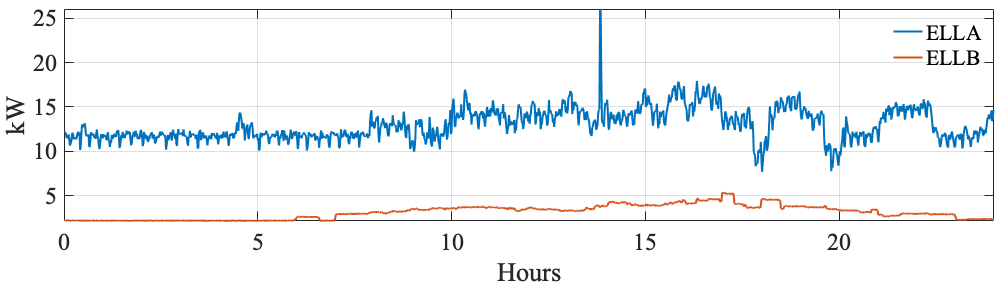}\label{fig:Load_day1}}\\
    \subfloat[Measured PV generation from PV1 and PV2+PV3 plants.]{\includegraphics[width=\linewidth]{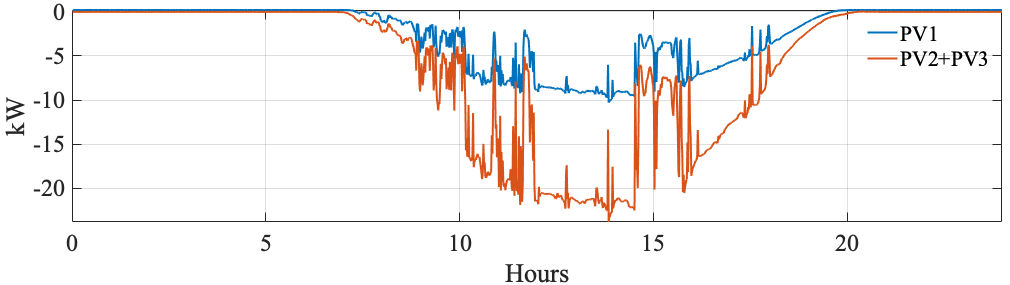}\label{fig:PV_day1}}
    \caption{Day-1 experimental dispatch tracking results.}
    \label{fig:GCP_day1}
\end{figure}

\subsection{Experimental results}
The experimental validation was performed for several days during the week. For the sake of brevity, {we present} results for two notable days of experiments. The days are characterised by different weather conditions and number of EV charging sessions. {The results are presented when there is no control of both BESS and EVCS, referred to as ``without control'' and when BESS and EVCS are both controlled, referred to as ``with control''.}
%%%%%%%%%
\subsubsection{Day 1}
It corresponds to a {weekday} and is characterised by cloudy irradiance {conditions}. The day shows  intermittent PV production and has several EV charging sessions.

\begin{figure}[!htbp]
    \centering
    \includegraphics[width=\linewidth]{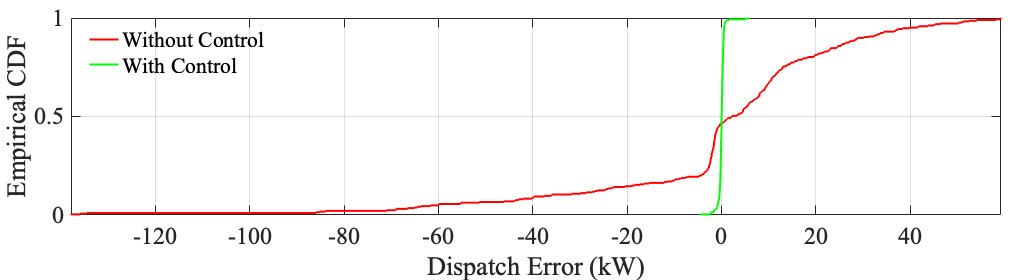}
    \caption{CDF plot of tracking dispatch error {without and with control} for day-1.}
    \label{fig:CDF_error_day1}
\end{figure}

Figure~\ref{fig:GCP_day1} and \ref{fig:CDF_error_day1} show the experimental results obtained for day-1. Figure~\ref{fig:GCP_dispatch_day1} shows the dispatch plan in shaded green (from day-ahead formulation described in Part-I) and realized power at the GCP with and without\footnote{Since each experiment day is unique with respect to the solar irradiance, number and energy demand of EV charging sessions, it is impossible to redo the same experiments in ``without control” mode. Therefore, we obtain the plot ``without control” by removing the contribution of the BESS and re-running the AC load flow with the rest of the injections. } control {(as shown in black and red color, respectively).}
Figure~\ref{fig:BESS_day1} shows the power injections and the SoC from the controllable battery BESS1. Figure \ref{fig:EVCS1_day1}-\ref{fig:EVCS2_day1} shows the EV demand (with control) and the EV SoC of the connected cars at the EVCS1 and EVCS2, respectively. In these figures, the target SoC is shown in red, and the SoC is shown in black. Figure ~\ref{fig:Load_day1} shows the uncontrollable demand (at nodes B20 and B21) and Fig.~\ref{fig:PV_day1} shows the PV generation (at nodes B14 and B16). 

One can observe from {Figure \ref{fig:GCP_dispatch_day1}}, that the dispatch plan is tracked with high fidelity thanks to the power injected from the controllable BESS and curtailment actions from EVCSs. From the plot, it can also be observed that the variation in the generation at the PV plants is well compensated by the battery storage. In Figure ~\ref{fig:BESS_day1}, the BESS SoC stays within the imposed SoC constraint of 20\% to 90\%. Figure~\ref{fig:EVCS1_day1}-\ref{fig:EVCS2_day1} shows the target SoC of the EVs, and in most of cases, EV users meet their target SoC. 
\begin{table}[!h]
\footnotesize
    \centering
    \caption{Tracking error statistics with different control schemes.}
    \begin{tabular}{|c|c|c|c|c|c|c|}
    \hline
     \bf{MPC} & \multicolumn{3}{|c|}{\bf{Day 1}} & \multicolumn{3}{|c|}{\bf{Day 2}} \\
     \hline 
            & RMSE & AEE & MAE & RMSE & AEE & MAE \\
            & (kW) & (kWh)  & (kW)  & (kW)  & (kWh)  & (kW)  \\
     \hline
      No Control    &   28.7    &   137.9     &  441.7    &  19.1   &  91.9    & 327.4\\ 
      \hline
      RT-MPC  &    0.7    &    5.9    &   8.5  &  0.5   &   2.9   & 1.5\\
      {(EVCS+BESS)} & & & & & & \\
      %Two-layer      &    89    &    1.5e3    &   332   &  85   &    1.5e3   & 322\\
      \hline
    \end{tabular}
    \label{tab:trackingerror}
\end{table}
%%% DAY2 %%%%%
\begin{figure}[!htbp]
    \centering
     \subfloat[Dispatch plan, and power at the GCP with and without control.]{\includegraphics[width=1\linewidth]{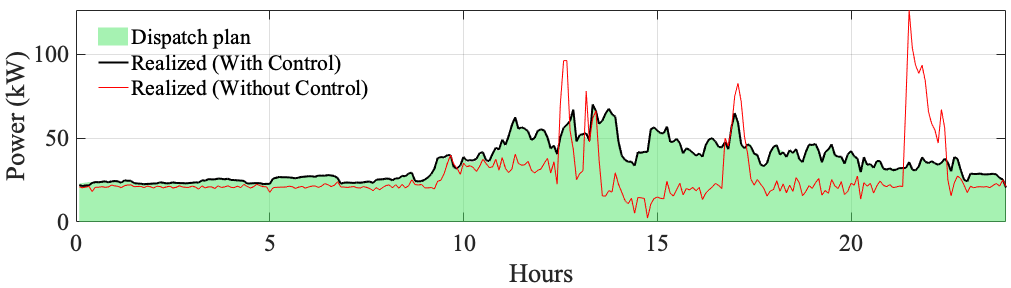}\label{fig:GCP_dispatch_day2}} \\
     \subfloat[Upper {panel}: active power regulation from the BESS1, and lower panel: {measured state-of-charge (SoC}).]{\includegraphics[width=\linewidth]{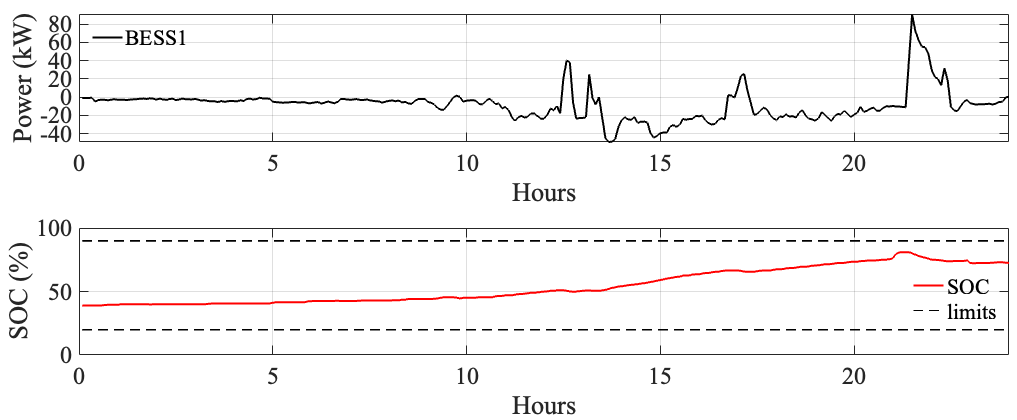}\label{fig:BESS_day2}}\\
     \subfloat[Upper {panel}: Controlled active power consumption by EVCS1, lower {panel}: EV SoC during the day along with the SoC target.]{\includegraphics[width=\linewidth]{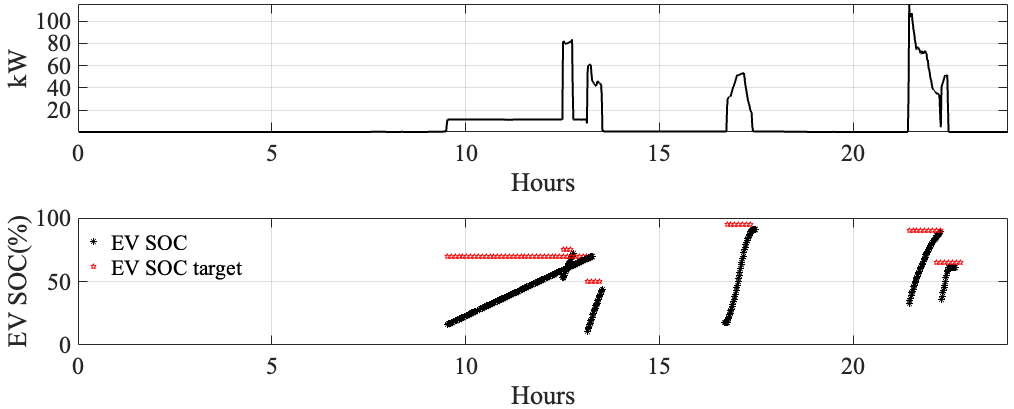}\label{fig:EVCS_day2}}\\
    \subfloat[Measured demand at ELLA and ELLB.]{\includegraphics[width=\linewidth]{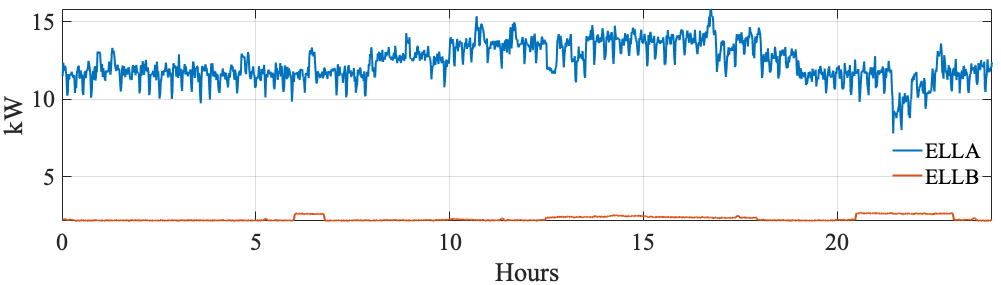}}\\
    \subfloat[Measured PV generation from PV1 and PV2+PV3 plants.]{\includegraphics[width=\linewidth]{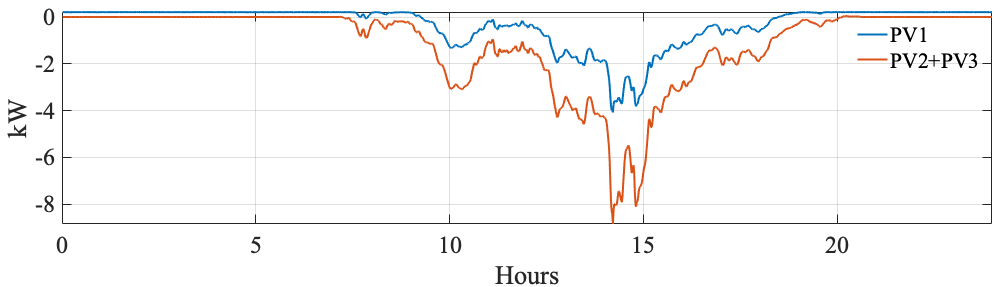}}
    \caption{Day-2 experimental dispatch tracking results.}
    \label{fig:GCP_day2}
\end{figure}
\begin{figure}[!htbp]
    \centering
    \includegraphics[width=\linewidth]{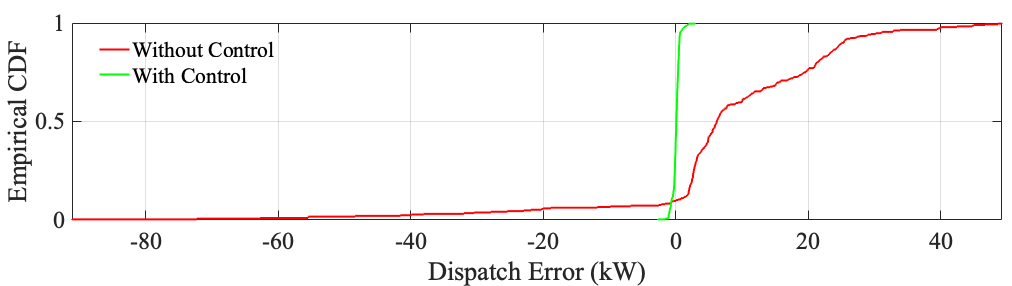}
    \caption{CDF plot of tracking dispatch error with and without error for day-2.}
    \label{fig:CDF_error_day2}
\end{figure}
%%%%
\begin{figure*}[!htbp]
    \centering
    \subfloat[Dispatch plan in shaded green area, power at the GCP with and without control.]{\includegraphics[width=0.7\linewidth]{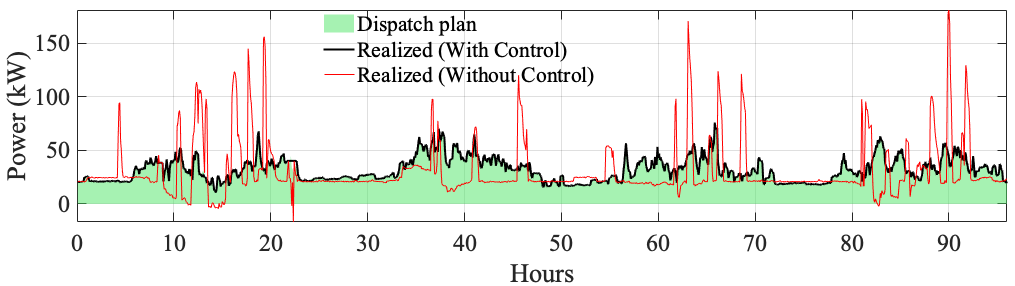}\label{fig:multiday_PCC}} \\
    \subfloat[Upper {panel}: BESS power injection and lower {panel}: measured BESS SoC evolution and limits.]{\includegraphics[width=0.7
\linewidth]{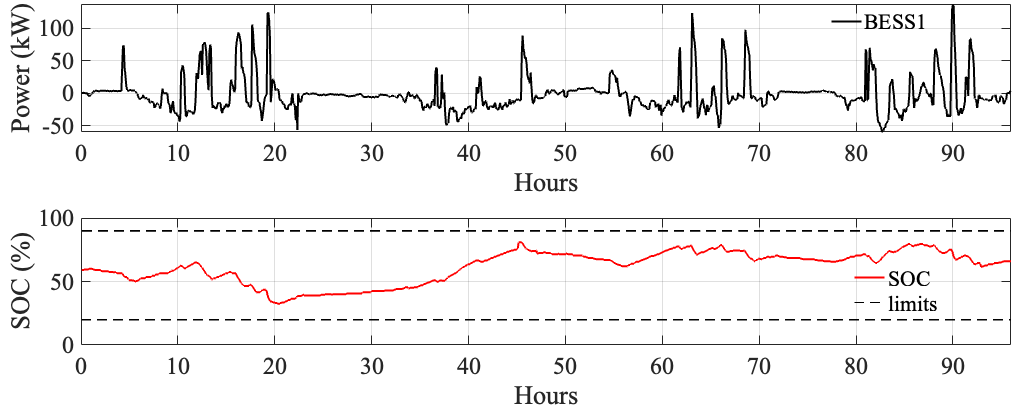}\label{fig:multidayBESS}} 
    \caption{Dispatch plan tracking results for multiple contiguous days of real-time operation.}
    \label{fig:multiday_results}
\end{figure*}

Table \ref{tab:trackingerror} shows different metrics to quantify the dispatch error with and without control. It shows the RMSE error, max absolute error (MAE), and Absolute Energy Error (AEE) of the dispatch over the day. AEE is defined as the absolute sum of the dispatch tracking error over the whole day. From the comparison, it is clear that the RT control manages to track with high accuracy, exhibiting low RMSE and MAE. The real-time control manages to reduce error metrics more than tenfold. Moreover, RMSE, AEE and MAE are reduced by 2.4\%, 4.2\% and 1.9\% respectively, compared to without control, proving the effectiveness of the proposed scheme.

We also show the cumulative distribution function (CDF) of the dispatch tracking error (averaged over the dispatch period of 5 minutes) with and without real-time control in Fig.~\ref{fig:CDF_error_day1}. As it can be observed, the dispatch error with control is always concentrated around zero. 
%%%%%%%%%%%
\subsubsection{Day 2}
It corresponds to a weekend day and is characterized by rainy {weather conditions}, so it exhibits low irradiance, leading to low PV generation and relatively low demand compared to the weekday. It also has less EV charging sessions.

Again, we show the active power realization at the GCP with and without control. It is shown in Figure~\ref{fig:GCP_dispatch_day2}. One can observe that the dispatch plan is tracked well, thanks to the power regulation provided by the controllable batteries, as shown in Figure~\ref{fig:BESS_day2}, and the curtailment action of EVCS1 as shown in Figure~\ref{fig:EVCS_day2}.
As this day corresponds to a rainy day (resulting in less PV generation), the peak power of the dispatch plan is higher than the one on day 1. On this day, there are no sessions on the EVCS2 (therefore, the graph is not shown). Indeed, that charging station belongs to the office's private space, which is unoccupied during the weekend. There are some charging sessions on the EVCS1 but less than on day 1 (due to weekend day). All sessions met their targets. 

Furthermore, {Figure}~\ref{fig:CDF_error_day2} shows the CDF of the dispatch error with and without control and it can be concluded that the real-time control achieves a very good accuracy in the dispatch tracking. The same can be observed by the metrics shown in Table~\ref{tab:trackingerror}, where it is observed that the RMSE, AEE and MAE are reduced to 2.6\%, 3.2\% and 0.5\%, respectively, compared to without control.
%%%%%
\subsubsection{Multi-day}
To demonstrate the effectiveness of the dispatching scheme, we ran the control of the BESS for four contiguous days shown in Fig.~\ref{fig:multiday_results}. Figure~\ref{fig:multiday_PCC} shows the dispatch plan and the measured GCP power with
and without the control scheme. In Fig.~\ref{fig:multidayBESS}, we show the SoC evolution of BESS1 during the four days. The power at the GCP follows the dispatch plan and keeps the BESS SoC within comfortable bounds so that dispatching is facilitated on the next day. 

\section{Conclusion}
\label{sec:conclusion}
Part-II of the paper presented the intra-day stage of the dispatching framework proposed in the Part-I paper. More specifically, it proposed and experimentally validated a real-time model predictive control for tracking a pre-defined power profile, the dispatch plan, at the grid connection point of an ADN by controlling flexible resources. The MPC control scheme is designed to control flexible resources such as EVCS and BESS to provide the mismatch between the dispatch plan and realization during the day. The dispatch plan is provided by the day-ahead scheduling formulation from Part-I. The MPC accounted for the grid constraints via a linearized power flow to obtain a tractable formulation. It considered the stochasticity of the uncontrollable generation and demand via short-term forecasting schemes.

The MPC scheme was experimentally validated on a real-life ADN at the EPFL's Distributed Electrical Systems Laboratory, which hosts two EVCSs and two BESSs as flexible resources and heterogeneous uncontrollable resources such as PV plants and office buildings. The RT-MPC was designed to run every 30 seconds with 5~minutes MPC horizon time to track the dispatch plan on a 5-minutes time resolution. 

The experimental validation carried out for several contiguous days proved the effectiveness of the MPC algorithm. The elaborated results presented for two distinct days showed good dispatch tracking accuracy. It has been observed that the metrics on dispatch tracking error such as root-mean-square-error, absolute energy error and maximum absolute error are reduced by factors ranging 38-41, 23-32, and 52-218, respectively by using the proposed MPC control scheme. 

\bibliographystyle{IEEEtran}
\bibliography{bibliography.bib}
\end{document}